\documentclass[epj]{svjour}

\usepackage{graphicx}
\usepackage{fancyhdr}
\def\makeheadbox{{%  remove EPJC header
 }}
\def\mytitle{My title} 
\def\myauthors{My name}  
\def\mytype{My type of session}
\def\mysession{My session}
\def\mytitle{Electroweak Contributions to Squark Pair Production} %Put your title here!
\def\myauthors{Sascha Bornhauser}    %Put your name here!
\def\mytype{Electroweak Contributions to Squark Pair Production}    
\def\mysession{Colliders - SUSY Phenomenology}
\newcommand{\ul} {\tilde u_L}
\newcommand{\ur} {\tilde u_R}
\newcommand{\dl} {\tilde d_L}
\newcommand{\dr} {\tilde d_R}
\newcommand{\ulb} {\bar{\tilde u}_L}
\newcommand{\urb} {\bar{\tilde u}_R}
\newcommand{\dlb} {\bar{\tilde d}_L}
\newcommand{\drb} {\bar{\tilde d}_R}

\newcommand{\gsim}{\raisebox{-0.13cm}{~\shortstack{$>$ \\[-0.07cm] $\sim$}}~}
\newcommand{\eqa} {\begin{eqnarray} }
\newcommand{\eqe} {\end{eqnarray}}
\newcommand{\beq} {\begin{equation}}
\newcommand{\eeq} {\end{equation}}
\usepackage{wrapfig}

\begin{document}
\title{\bf Electroweak Contributions to Squark Pair Production}
\author{
Sascha Bornhauser\thanks{\emph{Speaker}}, Manuel Drees, Herbi K. Dreiner
and Jong Soo Kim
% Sascha Bornhauser\inst{1}, Manuel Drees\inst{1}, Herbi K. Dreiner\inst{1}
% and Jong Soo Kim\inst{1}
}                     % Do not remove
\institute{\it Physikalisches Institut, Universit\"at Bonn, Nussallee 12, D53115
  Bonn,  Germany}
\date{}
\abstract{
  We compute electroweak contributions to the production of
  squark pairs at hadron colliders. These include the exchange of electroweak
  gauge bosons in the $s-$channel as well as electroweak gaugino exchange in
  the $t-$ and/or $u-$channel. In many cases these can interfere with the
  dominant QCD contributions. As a result, we find sizable contributions to
  the production of two $SU(2)$ doublet squarks. At the LHC, they amount to 10
  to 20\% for typical mSUGRA (or CMSSM) scenarios, but in more general
  scenarios they can vary between $-40$ and $+55\%$, depending on size and
  sign of the $SU(2)$ gaugino mass. The electroweak contribution to the total
  squark pair production rate at the LHC is about 3.5 times smaller.
} %end of abstract
\maketitle
\section{Introduction}
We compute the complete leading order electroweak contributions
to squark pair production at hadron colliders. Since in many cases
interference with QCD amplitudes is possible, this yields contributions of 
${\cal O}(\alpha_W^2)$ as well as of ${\cal O}(\alpha_S \alpha_W)$, where
$\alpha_W$ is a weak gauge coupling. We find that these change the total cross
section by only a few percent if at least one of the produced squarks is an
$SU(2)$ singlet. On the other hand, the cross section for the production of
two $SU(2)$ doublet squarks is changed by 10 to 20\% in typical mSUGRA
\cite{msugra} (or CMSSM) scenarios \cite{sps}; in scenarios without gaugino
mass unification \cite{code} the corrections can exceed
50\%. These new contributions peak
at small transverse momentum of the produced squark; they can therefore not be
subsumed in a constant ``$k-$factor''.

{This contribution is a short summary of Ref.\cite{ourp}, where 
detailed results for the leading--order parton--level 
squared matrix elements for the production of two (anti--) squarks from two (anti--)quarks in the initial 
state and additional explanations for the following numerical results are given. }

\section{Numerical Results}
{We focus on $pp$ collisions at the LHC operating at $\sqrt{s} = 14$ TeV
and squarks of the first and second generation, where mixing between $SU(2)$ doublets and
singlets can be neglected. }
Third generation squarks are produced dominantly
through gluon fusion or pure $s-$channel diagrams; the EW contributions to
these cross sections will therefore be very small. 
 \begin{table}
 \caption{Total cross sections at the LHC. 
All masses are in $10$GeV.
All cross sections are in pb. The last
  two columns show the ratio (QCD $+$ EW) $/$ QCD. We show results for the sum
  over all squark pairs (``tot''), as well as for the sum over all
   combinations of two $SU(2)$ doublet squarks (``LL''). }
\label{tab:1}       % Give a unique label
%For LaTeX tables use
\begin{tabular*}{8.45cm}{p{0.27cm}lp{0.29cm}lp{0.29cm}lp{0.29cm}lp{0.29cm}lp{0.29cm}lp{0.29cm}lp{0.29cm}lp{0.29cm}lp{0.29cm}l}
\hline\noalign{\smallskip}
& & & & \multicolumn{2}{c}{QCD} &
\multicolumn{2}{c}{QCD+EW} & \multicolumn{2}{c}{ratio} \\
 SPS & $m_0$ & $m_{1/2}$ & $m_{\tilde q}$ & tot & LL & tot & LL &
tot&LL\\
\noalign{\smallskip}\hline\noalign{\smallskip}
1a & 10 & 25 & 56 & 12.1 & 3.09 & 12.6 & 3.50 & 1.04 & 1.13 \\
1b & 20 & 40 & 87 & 1.57 & 0.42 & 1.66 & 0.50 & 1.06 & 1.19 \\
2 & 145 & 30 & 159 & 0.06 & 0.01 & 0.06 & 0.01 & 1.03 & 1.09
\\ 
3 & 9 & 40 & 85 & 1.74 & 0.46 & 1.83 & 0.55 & 1.06 & 1.19 \\
4 & 40 & 30 & 76 & 3.10 & 0.81 & 3.22 & 0.93 & 1.04 & 1.14 \\
5 & 15 & 30 & 67 & 5.42 & 1.41 & 5.66 & 1.62 & 1.04 & 1.15 \\
\noalign{\smallskip}\hline 
\end{tabular*} \vspace{-0.56cm}
 \end{table}
Table 1 shows results for the total squark pair production cross sections at
the LHC in six mSUGRA benchmark scenarios, taken from \cite{sps}. Here we sum
over all squarks and anti--squarks of the first and second generation; results
where both final state (anti--)squarks are $SU(2)$ doublets are shown
separately.  We only include contributions with (anti--)quarks in the initial
state, since the gluon fusion contribution obviously does not receive
electroweak contributions in leading order. We take equal factorization and renormalization scales,
$\mu_F = \mu_R = m_{\tilde q} / 2$; this choice leads to quite small NLO
corrections to the pure QCD contribution.

Not surprisingly, the cross sections fall quickly with increasing squark mass.
The partonic cross sections scale like $m^{-2}_{\tilde q}$, if the ratios of
sparticle masses are kept fixed and the running of $\alpha_s$ is ignored. In
addition, the pdf factors decrease quickly with increasing squark mass. There
is also some dependence on the gluino mass ($\simeq 2.5 m_{1/2}$), which
appears in $t-$ and $u-$channel propagators. Varying the ratio $m_{\tilde g} /
m_{\tilde q}$ between 0.5 and 1.2, which is the range covered by the scenarios
of Table~1, leads to 15 to 20\% variation of the QCD prediction.

Since QCD contributions dominate even after inclusion of the electroweak
diagrams, the overall behavior of the total cross sections does not change
much. These contributions are clearly more important for the production of
$SU(2)$ doublet squarks than for the total cross section summed over all final
states. This is not surprising: the cross sections for all other combinations
of squarks only receive electroweak contributions due to hypercharge
interactions, and the squared $SU(2)$ gauge coupling exceeds the squared
$U(1)_Y$ coupling by a factor $\cot^2 \theta_W \simeq 3.3$.

However, the weakness of the $U(1)_Y$ coupling by itself is not sufficient to
explain the small size of electroweak contributions to final states involving
at least one $SU(2)$ singlet $\mbox{\rm (anti--)squark}$. For example, we can
infer from the first line of Table 1 that in scenario SPS 1a, electroweak
contributions increase the cross section for the production of two $L-$type
squarks by 0.41 pb, whereas they only contribute 0.03 pb to all other squark
pair production channels combined. We also note that the importance of the EW
contributions seems to depend much more strongly on the ratio $m_{1/2} / m_0$
than the QCD prediction does. Finally, the EW contributions evidently become
more important for heavier squarks if the ratio $m_0 / m_{1/2}$ remains
roughly the same.

{In order to understand these features it is helpful to consider all 24 
different processes involving only (s)quarks and anti--(s)quarks from the first
generation which add up to the total cross section, cf. Table 2. These processes can be 
grouped into three categories having different weighting of the electroweak contributions 
with respect to the pure QCD cross section and positive or negative interference between 
 $s-$, $t-$ and $u-$channel diagrams, respectively. 

The first category consists of seven
reactions with interference between $t-$  and $u-$ channel diagrams, where in
all but the last case there are both strong and electroweak contributions from
both $t-$ and $u-$channel diagrams. The next class of seven processes allows
interference between $s-$ and $t-$channel diagrams. In the first four cases
there are both QCD and electroweak contributions to both the $t-$ and
$s-$channel, while in the last three cases only one QCD diagram contributes.
For the third class of ten processes, no interference between electroweak and
strong contributions is possible; two of these processes only proceed via
$s-$channel diagrams, whereas the remaining eight are pure $t-$channel
reactions.

The final total cross section depends on a 
complex interplay of the pdf's of the quarks, the mass and hypercharge of the squarks, 
a possible helicity flip of the exchanged $t-$ or $u-$channel fermion and whether the produced 
squarks are $SU(2)$ singlets or doublets and have to be a $S-$ or $P-$wave, respectively. 
For an extensive analysis we refer again to our paper \cite{ourp}. }
\begin{table}
\caption{The 24 different squark pair production processes involving first
  generation (s)quarks.}
\label{tab:2}       % Give a unique label
\begin{tabular}{llll}
\hline\noalign{\smallskip}
No. & Process & No. & Process   \\
\noalign{\smallskip}\hline\noalign{\smallskip}
1 & $u u \rightarrow \ul \ul$ & 13 & $d \bar d \rightarrow \ul \ulb$ \\
2 & $u u \rightarrow \ur \ur$ & 14 & $u \bar d \rightarrow \ul \dlb$ \\
3 & $u u \rightarrow \ul \ur$ & 15 & $u d \rightarrow \ul \dr$ \\
4 & $d d \rightarrow \dl \dl$ & 16 & $u d \rightarrow \ur \dl$ \\
5 & $d d \rightarrow \dr \dr$ & 17 & $u d \rightarrow \ur \dr$ \\
6 & $d d \rightarrow \dl \dr$ & 18 & $u \bar u \rightarrow \ul \urb$ \\
7 & $u d \rightarrow \ul \dl$ & 19 & $d \bar d \rightarrow \dl \drb$ \\
8 & $u \bar u \rightarrow \ul \ulb$ & 20 & $u \bar d \rightarrow \ul \drb$ \\
9 & $u \bar u \rightarrow \ur \urb$ & 21 & $u \bar d \rightarrow \ur \dlb$ \\
10 & $d \bar d \rightarrow \dl \dlb$ & 22 & $u \bar d \rightarrow \ur \drb$ \\
11 & $d \bar d \rightarrow \dr \drb$ & 23 & $u \bar u \rightarrow \dr \drb$ \\
12 & $u \bar u \rightarrow \dl \dlb$ & 24 & $d \bar d \rightarrow \ur \urb$ \\
\noalign{\smallskip}\hline
\end{tabular} \vspace*{-0.55cm}  % with the correct table height
\end{table}

{By looking at Table~1} we see that the relative importance of the electroweak
contributions increases with increasing gaugino to squark mass ratio.  This
can be explained from 
{the observation that the most important EW
contributions involve the interference of $t-$ and $u-$channel 
amplitudes \cite{ourp}. 
%As explained in Ref. \cite{ourp}, the  
The}
amplitudes for all processes
of this kind that receive contributions from $SU(2)$ interactions are
proportional to a gaugino mass. These contributions are therefore sensitive to
the ratio of gaugino and squark masses. In mSUGRA the relative importance of
the EW contributions becomes largely insensitive to $m_{1/2}$ (for fixed
squark mass) once $m_{1/2} \gsim m_0$. The physical squark masses are then
essentially independent of $m_0$, i.e. $m_{\tilde q} \propto m_{1/2}$, so that
the ratios of gaugino and squark masses become independent of $m_{1/2}$. 

\begin{figure}[h!] 
\begin{center}
%\rotatebox{270}{\includegraphics[width=7.50cm,height=4.2cm]{finalptss.eps}}
\rotatebox{270}{\includegraphics[width=4.7cm,height=8.5cm]{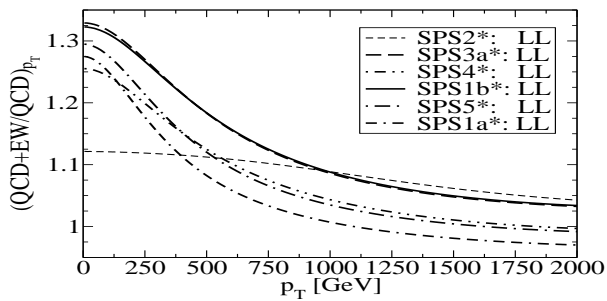}}
\caption{The ratio of QCD$+$EW to pure QCD predictions  
as a function of the squark
  transverse momentum. \vspace{-0.75cm} }
\label{fig1}
\end{center}
\end{figure}

Finally, Table~1 also shows that the electroweak contributions become
relatively more important with increasing squark mass scale, although for
scenario SPS 2 this effect is over--compensated by the small ratio $m_{1/2} /
 m_0$. 
{The reason for this is the different relative importance (decrease or increase) of the 24 processes for the 
three categories due to the suppression of the PDFs at the required larger values of Bjorken$-x$ and/or the threshold factor $\beta$ (which is the squark center--of--mass [cms] velocity).}
{We pointed out in Ref. \cite{ourp} }that the dominant EW contributions come from the
interference of $t-$ and $u-$channel diagrams with QCD diagrams. Since in
mSUGRA the electroweak gauginos are about three and six times lighter than the
gluino, one expects the EW contributions to be most prominent for small
transverse momenta of the produced squarks. This is borne out by
Fig.~\ref{fig1}, which shows the ratio of the tree--level differential cross
section with and without EW contributions. Here, and in the subsequent
figures, we concentrate on the production of two $SU(2)$ doublet
(anti--) squarks, where the EW contributions are largest.
The observed behavior can be understood from the interplay of several effects.
For simplicity assuming equal squark masses in the final state, the relation
between the partonic cms energy and squark transverse momentum can be written
as
\beq \label{ept}
\hat{s} = 4 \left( m^2_{\tilde q} + \frac{p_T^2 } {\sin^2 \theta} \right)\, ,
\eeq
where $\theta$ is the cms scattering angle. The parton flux in the initial
state is largest for smallest $\hat{s}$. Eq.(\ref{ept}) then shows that
configurations where $\sin^2\theta$ is maximal, i.e. where $\cos\theta$ is
small, are preferred if $p_T$ is sizable.

On the other hand, the denominators of the $t-$ channel propagators can be
written as
\beq \label{eprop1}
\hat{t} - M^2_{\tilde V} = m^2_{\tilde q} -\frac{\hat{s}} {2} ( 1 - \beta
\cos\theta)  - M^2_{\tilde V} \, , 
\eeq
where $M_{\tilde V}$ is the mass of the exchanged gaugino; the expression for
$u-$channel propagators can be obtained by $\cos\theta \rightarrow
-\cos\theta$. These propagators therefore prefer large $\beta |\cos\theta|$;
however, $t-$ and $u-$channel propagators prefer different signs of
$\cos\theta$. 

% {\it As explained in Ref.\cite{ourp} }
The dominant EW contributions are due to the
interference between $t-$ and $u-$channel diagrams. 
These cross sections are proportional to a single power of the threshold
factor $\beta$. The steeply falling pdf's imply that these processes therefore
prefer rather small values of $\beta$ even for small $p_T$. As a first
approximation 
we can therefore ignore terms $\propto \beta \cos\theta$ in the
propagators. The ratio of EW and QCD $t-$ or $u-$channel propagators then
becomes
\beq\label{proprat}
\frac {\rm EW} {\rm QCD} = \frac {\hat{s}/2 - m^2_{\tilde q} + M^2_{\tilde g}
} {\hat{s}/2 - m^2_{\tilde q} + M^2_{\widetilde W}} \simeq
\frac {2 p_T^2 + m^2_{\tilde q} + M^2_{\tilde g}}
{2 p_T^2 + m^2_{\tilde q} + M^2_{\widetilde W}}\,,
\eeq
where $M_{\widetilde W}$ is the mass of the relevant chargino or neutralino.
Most of the mSUGRA scenarios of Table~1 and Fig.~\ref{fig1} have $m^2_{\tilde
  q} \sim M^2_{\tilde g} \gg M^2_{\widetilde W}$. Eq.(\ref{proprat}) shows
that the interference term will then be enhanced by a factor $\sim 2$ at small
$p_T$.  

However, this enhancement disappears for $m^2_{\tilde q} \gg
M^2_{\tilde g}$, as in SPS 2.
Eq.(\ref{proprat}) shows that the propagator enhancement of the EW
 contributions also disappears once $2 p_T^2 \gg m^2_{\tilde q}$. However, at
 large $p_T$ 
{one has two kinds of competing processes \cite{ourp}: On the one hand constructive 
interference, where EW contributions enhance the cross section but are 
suppressed by an extra factor of $p_T^{-2}$ due to a necessary helicity flip. 
On the other hand destructive interference and without a helicity flip which are 
suppressed by more quickly falling PDFs. }
Eq.(\ref{ept}) shows
that the latter suppression will be more relevant for larger squark masses.
Indeed, at large $p_T$ we observe the largest (or least negative) EW
contributions for scenarios with heaviest squarks. However, even in scenario
SPS1a, which has the smallest squark masses, EW contributions only suppress
the cross section by $\sim 3\%$ at large $p_T$.

\begin{figure}[h!] 
\begin{center}
%\rotatebox{0}{\includegraphics[width=7.5cm,height=4.2cm]{finalslopess.eps} }
\rotatebox{270}{\includegraphics[width=4.4cm,height=8.2cm]{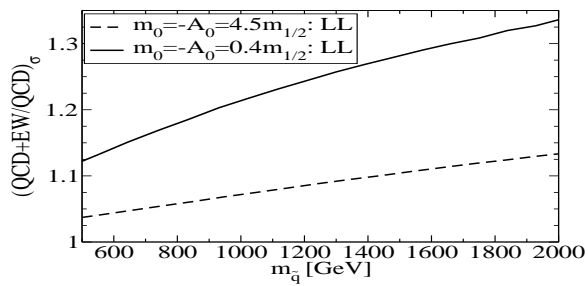} }
\caption{The ratio of QCD$+$EW to pure QCD predictions as a function of the squark
  mass. \vspace{-0.75cm}}
\label{fig2}
\end{center}
\end{figure}

We saw in Table~1 that the EW contributions tend to become more important with
increasing squark mass scale. This is further explored in Fig.~\ref{fig2},
which shows the ratio of the total cross section for the production of $SU(2)$
doublet squarks with and without EW contributions as function of the average
doublet squark mass. These curves have been generated by keeping the ratios of
the dimensionful mSUGRA input parameters $m_0, \, m_{1/2}$ and $A_0$ fixed,
but varying the overall mass scale; this corresponds to the ``benchmark
slopes'' of ref.\cite{sps}. We see that in a scenario with relatively large
gaugino masses, as in SPS 1a (upper curve), the EW contribution can increase
the cross section by more than 30\% for $m_{\tilde q} = 2$ TeV. 
A scenario with $m_0 = -A_0 = 4.5 m_{1/2}$ (lower curve) shows the same
trend; however, as noted earlier, the total EW contribution is much smaller in
this case, only reaching 13\% for $m_{\tilde q} = 2$ TeV.
The results of
this Figure can therefore not be entirely due to the change of the relative
weights of the various processes.
%, as described in {\it Ref.\cite{ourp}.}
On top of that, the importance of the EW contributions to single processes
increases with increasing squark masses. This can be understood from the
behavior of the $t-$ and $u-$channel propagators. Smaller squark masses allow
larger values of $\beta$. The regions of phase space with large $|\cos
\theta|$ will then favor the squared $t-$ or $u-$channel propagators of pure
QCD contributions over the product of one $t-$ and one $u-$channel propagator
of the interference terms. This implies that increasing $m_{\tilde q}$ will
increase the relative importance of the interference terms relative to the
squared $t-$ and $u-$channel diagrams. This reduces the pure QCD contribution,
where the interference is destructive due to the negative color factor, see
e.g. Eq. (4) {of \cite{ourp}}, and enhances the importance of the EW contributions.
Comparison of the two curves in Fig.~\ref{fig2} reinforces the importance of
the gaugino masses. So far we have considered sparticle spectra generated with
mSUGRA boundary conditions. In particular, this implies that $SU(2)$ and
$U(1)_Y$ gauginos are much lighter than gluinos. Since the dominant EW
contributions are proportional to the product of
the gluino mass with the mass of an electroweak gaugino{ \cite{ourp}}, we expect that these
contributions are sensitive to the assumed ratio of gaugino masses.
\begin{figure}[h!] 
\begin{center}
%\rotatebox{0}{\includegraphics[width=7.5cm,height=3.65cm]{finalXsm2msquarkss.eps} }
\rotatebox{270}{\includegraphics[width=3.65cm,height=8.2cm]{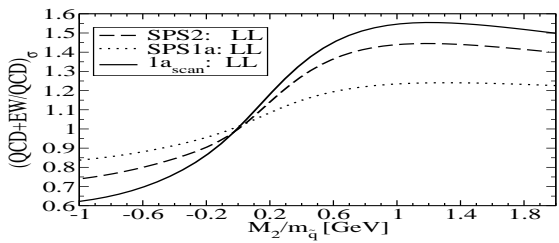} }
\caption{The ratio of QCD$+$EW to pure QCD predictions 
as a function of the ratio of the $SU(2)$ gaugino mass parameter $M_2$ and the squark mass. The solid and
dotted curves are both based on scenario SPS 1a of Table~1, but for the
solid curve all soft breaking masses have been scaled up to achieve a squark
mass of 2 TeV. \vspace{-0.75cm}}
\label{fig3}
\end{center}
\end{figure}
This is demonstrated in Fig.~\ref{fig3}, where we vary the $SU(2)$ gaugino
mass $M_2$ at the weak scale, keeping all other parameters fixed. We see that
the electroweak contributions become maximal if $M_2 \simeq m_{\tilde
  q}$. This can be understood from the observation that this choice maximizes $M_2 /
|\hat{t} - M_2^2|$, see Eq.(\ref{eprop1}). In a scenario with $m_{\tilde g}
\simeq m_{\tilde q}$ and large squark mass (solid curve), this can lead to EW
contributions in excess of 50\%. In scenario SPS 2 (dashed curve) the
contributions remain somewhat smaller, partly because of the reduced squark
mass, and partly because the lower gluino mass reduces the importance of the
interference terms. Not surprisingly, taking $m_{\tilde g} \simeq m_{\tilde
  q}$ also maximizes the size of those pure QCD contributions that require a
helicity flip.

Finally, in scenario SPS 1a with its relatively light squarks (dotted curve) the EW contribution
never goes much beyond 20\%. {In this case
%,  as described in Ref.\cite{ourp}, 
processes with negative EW contributions contribute significantly.} 
Since these processes do not require a helicity flip, the absolute
size of the EW contributions decreases monotonically with increasing $|M_2|$.
As a result, the dotted curve reaches its maximum for somewhat larger values
of $M_2$; moreover, the maximum is less pronounced.

In Fig.~\ref{fig3} we show results as function of the weak scale soft breaking
parameter $M_2$, normalized to the squark mass. This parameter can be
negative. If we keep the sign of the gluino mass parameter positive, the sign
of the $t-u$ interference terms, which require a helicity flip, will
change.We see that taking
$M_2$ large and negative will lead to cross sections that are significantly
reduced from the pure QCD contribution. The relative size of the EW
contributions is slightly smaller than that for positive $M_2$. This is partly
because we did not change the sign of the $U(1)_Y$ gaugino mass, keeping the
corresponding contribution positive (but very small). 

{EW contributions will be much smaller if at least one
(anti--)squark in the final state is an $SU(2)$ singlet \cite{ourp}}. However, it might
well be possible to experimentally separate these different classes of final
states. At least for $m_{\tilde g} \gsim m_{\tilde q} > |M_2|, |M_1|$ the
production of two doublet squarks leads to significantly different final
states than that of singlet squarks{\cite{bbkt}.} 
\section{Summary and Conclusions}

{We} analyzed electroweak (EW) contributions to the production of
two squarks or anti--squarks at the LHC. 
Not surprisingly, corrections due to $SU(2)$ interactions are more important than
those from $U(1)_Y$ interactions. In both cases the dominant effect is from
the interference of electroweak and QCD interactions. 

{In conclusion
the physical significance of our results are:} 
\begin{itemize}
  
\item The EW contributions can change the total cross section significantly.
  Focusing on the production of two $SU(2)$ doublet ($L-$type) squarks, we
  found the contributions with interference between $t-$ and $u-$channel
  diagrams to be dominant. For squark masses near the discovery reach of the
  LHC, EW effects can reduce or enhance the total cross section by more than a
  factor 1.5, if the absolute value of the $SU(2)$ gaugino soft breaking mass
  is near $m_{\tilde q}$; even in scenarios with gaugino mass unification the
  EW contribution can still change the cross section for the production of two
  $SU(2)$ doublet squarks by more than a factor 1.3. Recall that $SU(2)$
  doublet squarks often lead to different final states than singlet squarks
  do, allowing to distinguish these modes experimentally.
  
\item The EW contributions might give a new, independent handle on the gaugino
  mass parameters. In particular, we just saw that they are sensitive to
  relative {\em signs} between gaugino mass parameters, which might be
  difficult to determine using kinematical distributions only. For example, in
  anomaly--mediated supersymmetry breaking \cite{amsb} the products of
  electroweak and QCD gaugino masses are negative. In order to realize this
  potential, both the experimental and the theoretical uncertainties should be
  reduced to the 10\% level. 
This is certainly challenging, but should eventually be possible if squarks are not too heavy.

\end{itemize}

\end{document}